\documentclass[10pt, journal]{IEEEtran}

\usepackage{amsmath}
\usepackage{amsfonts} 
\ifCLASSOPTIONcompsoc
  \usepackage[nocompress]{cite}
\else
  \usepackage{cite}
\fi
\usepackage{bm} 
\usepackage{upgreek} 
\usepackage{psfrag} 
\usepackage[caption=false,font=footnotesize]{subfig} 
\usepackage{algpseudocode} 
\usepackage{mathtools} 
\usepackage{tikz}
\usepackage{booktabs} 
\usepackage[flushleft]{threeparttable} 
\usepackage{paralist} 

\DeclareMathOperator*{\argmin}{arg\,min}
\DeclareMathOperator{\OMP}{OMP}

\renewcommand{\imath}{\textrm{j}}

\newcommand{\vv}[1]{{\boldsymbol{\mathbf{#1}}}} 
\DeclarePairedDelimiter{\norm}{\lVert}{\rVert}

\title{Fast Rotational Sparse Coding}

\author{
Michael~T.~McCann,~\IEEEmembership{Member,~IEEE}
Vincent~Andrearczyk,
Michael~Unser,~\IEEEmembership{Fellow,~IEEE}
Adrien~Depeursinge,~\IEEEmembership{Member,~IEEE}%
\IEEEcompsocitemizethanks{
\IEEEcompsocthanksitem This project has received funding from the European Research Council (ERC) under the European Union's Horizon 2020 research and innovation program (GA No. 692726 GlobalBioIm).
It was also partly supported by the Swiss National Science Foundation under grants PZ00P2\_154891 and 205320\_179069.
\IEEEcompsocthanksitem Michael McCann is with the Dept. of Computational Mathematics, Science and Engineering, Michigan State University, East Lansing, MI (email: mccann13@msu.edu).
\IEEEcompsocthanksitem Michael Unser is with the Biomedical Imaging Group, EPFL, Lausanne, Switzerland.
\IEEEcompsocthanksitem Vincent Andrearczyk and Adrien Depeursinge are with the MedGIFT group, Institute of Information Systems, University of Applied Sciences Western Switzerland (HES-SO), Sierre 3960, Switzerland.
}}

\begin{document}

\def\arraystretch{1.2} 

\newlength{\figwidth}
\setlength{\figwidth}{252.0pt}

\maketitle

\begin{abstract}
  We propose an algorithm for rotational sparse coding
  along with an efficient implementation using steerability.
  Sparse coding (also called \textit{dictionary learning}) is an important technique
  in image processing, useful in inverse problems, compression, and analysis;
  however, the usual formulation fails to capture an important aspect of the structure of images:
  images are formed from building blocks, e.g., edges, lines, or points,
  that appear at different locations, orientations, and scales.
  The sparse coding problem can be reformulated to explicitly account for these transforms,
  at the cost of increased computation.
  In this work, we propose an algorithm for a rotational version of sparse coding
  that is based on K-SVD with additional rotation operations.
  We then propose a method to accelerate these rotations by 
  learning the dictionary in a steerable basis.
  Our experiments on patch coding and texture classification
  demonstrate that the proposed algorithm is fast enough for practical use
  and compares favorably to standard sparse coding.
\end{abstract}
\begin{IEEEkeywords}
steerable filters, sparse coding, dictionary learning, rotation invariance.
\end{IEEEkeywords}

\section{Introduction}
Sparse coding, also called \textit{sparse dictionary learning},
has had an immense impact in imaging, underpinning state-of-the-art methods in
inverse problems (denoising, reconstruction, inpainting), compression, and analysis (classification, detection)~\cite{ToF2011}.
The aim of sparse coding is to build a \textit{dictionary} comprised of elements called \textit{atoms} such that
each member of a dataset of interest can be approximated using a combination of a small number of atoms.
The underlying idea is that while data is often apparently high-dimensional (e.g., an image patch, $\vv{x} \in \mathbb{R}^{11\times 11}$),
we may be able to uncover a simple (i.e., sparse or low-dimensional) representation.
And, further, that the representation can be made even sparser by adapting the dictionary to the data, rather than by using a fixed dictionary (e.g., Fourier or wavelets).

The standard approach to sparse coding uses a linear model for the signal, $\vv{x} = \vv{D}\vv{a}$, where $\vv{x}$ is, e.g., a vectorized $N$ by $N$ image patch.
Starting from a collection of $P$ examples, $\vv{x}_p \in \mathbb{R}^{N^2}$,
it seeks a dictionary of $M$ atoms, $\vv{D} \in \mathbb{R}^{N^2 \times M}$,
and sparse codes,  $\vv{a}_1$, $\vv{a}_2$, \dots, $\vv{a}_P$,
such that each example is well-represented by just $K$ atoms,

\begin{equation}
  \label{eq:SC}
  \argmin_{\vv{D}, \{\vv{a}_p\}}
  \sum_{p=1}^P
  \norm{  \vv{x}_p - \vv{D} \vv{a}_p   }^2
  \quad \text{s.t.} \quad \norm{ \vv{a}_p }_0 \le K.
\end{equation}
No constraints are placed on the atoms of the dictionary:
they need not be orthogonal and they may not span the domain of the $\vv{x}$'s.
There are several variations on this formulation;
e.g., the $\ell_0$ norm can be replaced by an $\ell_1$ norm,
and the sparsity constraint can be reformulated as a regularization on the $\vv{a}$'s.
Notable algorithms for this type of dictionary learning include
the method of optimal directions (MOD)~\cite{EAH1999},
K-singular value decomposition (K-SVD)~\cite{AEB2006},
online dictionary learning~\cite{MBP2009},
and
recursive least squares dictionary learning~\cite{skretting_recursive_2010}.
Formulations like \eqref{eq:SC} are components of algorithms in many areas of image processing,
including
medical image reconstruction~\cite{ravishankar_mr_2011,xu_low_2012,chen_artifact_2014},
multispectral change detection~\cite{lu_joint_2017},
face recognition~\cite{lu_simultaneous_2017},
and histopathology image classification~\cite{vu_histopathological_2016}.

\begin{figure}[!htbp]
  \subfloat[true atom]{\includegraphics[trim = -20 47 27 0, clip, width=.31\figwidth]{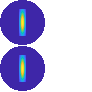}}\hfill
  \subfloat[representative image patches]{\includegraphics[trim = 0 47 0 0, clip, width=.65\figwidth]{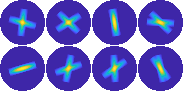}}\\
  \subfloat[standard atom]{\includegraphics[trim = -20 47 27 0, clip, width=.31\figwidth]{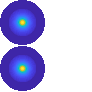}}\hfill
  \subfloat[K-SVD coding, MSE 15.31]{\includegraphics[trim = 0 47 0 0, clip,width=.65\figwidth]{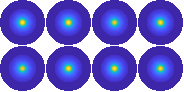}}\\
  \subfloat[standard atoms]{\includegraphics[angle=90,trim = 0 0 47 0, clip, width =.32\figwidth]{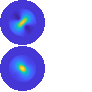}}\hfill
  \subfloat[K-SVD coding, MSE 10.35]{\includegraphics[trim = 0 47 0 0, clip,width=.65\figwidth]{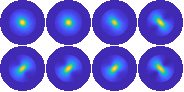}}\\
  \subfloat[steerable atom]{\includegraphics[trim = -20 47 27 0, clip, width=.31\figwidth]{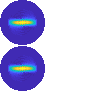}}\hfill
  \subfloat[R-K-SVD coding, MSE 3.64]{\includegraphics[trim = 0 47 0 0, clip,width=.65\figwidth]{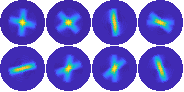}}
  \caption{Toy experiment involving coding patches ($P=1000$) that are formed as the sum of two rotated versions of a single atom (a-b).
Standard sparse coding (K-SVD, $M=1,2$, $K=2$) yields dictionaries from angular averaging of the patches (c, e).
Rotational sparse coding  (R-K-SVD, $M=1$, $K=2$) successfully recovers the true atom (e).
The resulting patch representations (d, f and h) are better in the rotational case (h), even when $M=1$.}
  \label{fig:toyCoding}
\end{figure}

When considering data in the form of image patches,
it is easy to find cases where standard linear sparse coding fails to capture
the underlying structure of the data.
As an example, consider a dataset consisting of crosses
formed from two random rotations of a line template (or \textit{true atom}, Figure~\ref{fig:toyCoding}(a) and (b)).
We would ideally code the dataset by storing the template
and two numbers per patch specifying the orientations of the two lines.
However, learning a dictionary according to \eqref{eq:SC} will not recover the template,
but rather blurry, circular harmonic-like atoms (Figure~\ref{fig:toyCoding}(e)).

It is natural to think of images being built from elements
that appear at various positions, orientations, and scales
(or, more generally, lie on the orbit of some group of transforms).
As the previous example illustrates,
when a linear sparse coding model such as \eqref{eq:SC} is used to code these data,
\begin{inparaenum}
\item  the resulting codes will not be as sparse as we would expect (or equivalently, if codes are forced to be sparse, the coding error will be intolerably high); and
  \item the span of the resulting dictionary will be much larger than we expect.
\end{inparaenum}
Point one is detrimental in compression applications, where coding error and sparsity are traded off against each other.
Point two is detrimental in classification applications, where the goal is to find a discriminative dictionary,
i.e. one that is well-adapted to the data it was learned from.

Several approaches have been proposed to address these shortcomings of sparse coding,
often called \textit{translation-invariant} or \emph{convolutional} (when they address translation),
\textit{rotation-invariant} (when they address rotation),
or \textit{transform-invariant} (when they address the problem more generally).
One approach is to first register the patches,
e.g., in terms of scale and rotation as in \cite{chen_plane_2013},
before performing standard sparse coding.
Such an approach is insufficient in cases like Figure~\ref{fig:toyCoding},
where patches are made of multiple atoms at different rotations and therefore cannot be aligned.
Another class of approaches involves reformulating the coding model itself.
We provide a brief summary of these approaches here for context;
for a recent review, see \cite{garcia-cardona_convolutional_2018}.
To handle patch translations, \cite{BEL2013} uses a convolutional model $\vv{x} = \sum_{m=1}^M \vv{d}_m \ast \vv{a}_{m}$,
where $\vv{x}$ and $\vv{a}$ represent entire images rather than patches.
Alternatively, the dictionary can be augmented with all its circular shifts $\vv{T}_r$, $\vv{x} = \sum_{r=1}^R \vv{T}_r \vv{D} \vv{a}_{r}$~\cite{JVL2006}.
The same idea can be used with arbitrary transformations in the place of $\vv{T}_r$, e.g., rotations or scaling~\cite{wersing_sparse_2003},
and the convolutional model and transform model can be combined,  $\vv{x} = \sum_{r=1}^R \sum_{m=1}^M (\vv{T}_r \vv{d}_m) \ast \vv{a}_{r,m}$~\cite{morup_transformation_2011}.
Another formulation is the union of submodules~\cite{aeron_group_2015},
where each patch is represented by a linear combination of the other patches
and all of their transforms.
These transform-invariant approaches continue to find
application in image processing,
e.g., in
superresolution~\cite{gu_convolutional_2015},
denoising~\cite{wohlberg_convolutional_2016},
and
multimodal imaging~\cite{degraux_online_2017}.

The main downside to these models is the computational cost:
\cite{BEL2013} reports minutes to code thousands of 50 by 50 patches,
\cite{wersing_sparse_2003} reports 1 to 2 days for 2000 4 by 4 patches,
and \cite{morup_transformation_2011} reports minutes for a 256 by 256 image.
In the latter two cases, the computational bottleneck arises from repeatedly applying the $\vv{T}_r$ operators.

Our approach to reducing this computational cost
involves the use of steerability.
Steerability refers to expressing the rotation of an image (or, equivalently, a filter or image patch)
by a suitable linear combination of a small set of images.
These ideas underlie the classic steerable image pyramid~\cite{simoncelli_shiftable_1992},
and, more recently, have been used to define a plethora of steerable wavelet transforms~\cite{Unser_Unifying_2013}.
Our use case here is unique in that we seek a discrete steerable basis,
i.e., a matrix that diagonalizes the operation of image rotation,
with the additional constraint that the matrix should be orthogonal.
A similar discrete design was considered in \cite{uenohara_optimal_1998},
but with a focus on coding a single rotated patch, rather than finding a basis for a set.

In this work, we propose a fast algorithm for performing rotational sparse coding.
We use a transform formulation for the learning problem, along the lines of \cite{wersing_sparse_2003,morup_transformation_2011};
but, we accelerate the learning by using steerability.
Specifically, we represent the dictionary atoms and input patches in a discrete steerable basis,
thereby diagonalizing the rotation operator.
For purposes of rough comparison, this representation allows
performing rotational sparse coding on 10,000 11 by 11 patches in 45 seconds.
Forcing the dictionary atoms to conform to a specific parametric form
has been previously explored---e.g. in \cite{chabiron_fast_2014},
where atoms are represented as convolutions of sparse filters---%
but we are not aware of a previous use of steerability in this context.

The remainder of the paper is organized as follows.
In Section~\ref{sec:methods}, we formulate the rotational sparse coding problem,
present an algorithm for solving it,
and describe our approach to accelerating this algorithm using a steerable basis.
In Section~\ref{sec:experiments}, we present experiments and results on patch coding, texture classification, and image rotation.
We conclude with a brief discussion in Section~\ref{sec:discussion}.

\section{Methods}
\label{sec:methods}
\subsection{Notation}
Throughout the paper, we use the following conventions.
Italic symbols represent scalars and functions, while bold, upright symbols represent (finite-length) vectors and matrices.

\subsection{Problem Statement}
We formulate the rotational sparse coding problem,
where patches are coded by a sparse linear combination of rotated dictionary atoms,
\begin{equation}
  \label{eq:rot-SC}
  \argmin_{\vv{D}, \{\vv{a}_{p,r}\}}
  \sum_{p=1}^P 
  \norm*{ \vv{x}_p - \sum_{r=0}^{R-1} \vv{R}_r \vv{D} \vv{a}_{p,r} }^2
   \quad \text{s.t.} \quad \sum_{r=0}^{R-1} \norm{ \vv{a}_{p,r} }_0 \le K,
\end{equation}
where $r$ indexes discretized rotation and $\vv{R}_r$ is an $N^2$ by $N^2$ matrix that performs a spatial rotation by an angle $\frac{2\pi r}{R}$ on a vectorized image patch.

This formulation is invariant to rotations in the sense
that applying random rotations to the input patches
does not change the coding error.
This is because
\begin{equation}
  \begin{split}
    \norm*{ \vv{R}_{r_0}\vv{x} - \sum_{r=0}^{R-1} \vv{R}_r \vv{D} \vv{a}_{r} }
    =  \norm*{ \vv{R}_{r_0}\vv{x} -\vv{R}_{r_0} \sum_{r=0}^{R-1} \vv{R}_{r-r_0} \vv{D} \vv{a}_{r} }\\
    =  \norm*{ \vv{x} - \sum_{r=0}^{R-1} \vv{R}_{r-r_0} \vv{D} \vv{a}_{r} }
     =  \norm*{ \vv{x} - \sum_{r=0}^{R-1} \vv{R}_{r} \vv{D} \vv{a}_{r+r_0} }
  \end{split}
\end{equation}
(where arithmetic on $r$ is taken modulo $R$),
which is just a relabeling of the $\{\vv{a}_{p,r}\}$ in \eqref{eq:rot-SC}.
From another perspective,
we can call the formulation rotation covariant
(also called \textit{invariant} in the signal processing sense),
because rotating a patch,  $\vv{x}_{p'} = \vv{R}_{r_0} \vv{x}_p$,
causes a commensurate rotation in its sparse code,  $\vv{a}_{p',r} = \vv{a}_{p, (r+r_0)}$.
We prefer the term \textit{rotational} because it emphasizes the presence of
the rotation matrix in the patch coding model
and parallels the use of \textit{convolutional} in \cite{BEL2013}.

Comparing the standard \eqref{eq:SC} and rotational \eqref{eq:rot-SC} formulations,
we see that, in both cases, each patch is coded by $K$ atoms from the dictionary.
In the standard formulation, each $\vv{a}_p$ is a $M$ by 1, $K$-sparse vector that defines which $K$ atoms in the dictionary are used and what is their weight;
in the rotational formulation, the $\vv{a}_{p,r}$ for each patch can be concatenated into a $MR$ by 1, $K$-sparse vector that defines which $K$ atoms are used, how they are rotated, and what is their weight.
In the standard formulation, it is nonsense to set $K > M$ because $\vv{a}_p$ cannot have more nonzero entries than it has total entries;
in the rotational formulation, setting $K > M$ means that some atoms will be used multiple times in different orientations, e.g. see Figure~\ref{fig:toyCoding}.
Moreover, even when $K \le M$, the same atom may be used multiple times to code a given patch if it is advantageous to do so.

\subsection{Rotational K-SVD}
In order to solve the optimization problem \eqref{eq:rot-SC}, we use a procedure that alternates between updating the sparse codes, $\vv{a}$, and the dictionary, $\vv{D}$.
This mimics K-SVD\cite{AEB2006}, but with critical changes to correctly handle the rotations;
we therefore call it rotational K-SVD (R-K-SVD).

We now describe the algorithm in detail;
see Figure~\ref{fig:code} for pseudocode.
The input for the algorithm is a set of image patches, $\{\vv{x}_p\}$, the dictionary size $M$, and the sparsity $K$.
The number of discretization levels for rotation, $R$, is a parameter that increases the quality, but also runtime of the algorithm, typical values range from ten to one hundred.
The dictionary is initialized with a random set of unit-norm vectors.
The main loop begins with all rotated versions of the dictionary atoms being computed
and concatenated to form the augmented dictionary, $\vv{\mathcal{D}}$.
This augmented dictionary is used to find $K$-sparse codes for the patches using orthogonal matching pursuit (OMP)~\cite{pati_orthogonal_1993},
which solves the sparse coding problem (equivalent to \eqref{eq:SC} with $\vv{D}$ fixed) greedily.
The resulting sparse codes are $MR$ by 1 vectors for each image patch.
These can be reshaped into a set of $R$ $M$ by 1 vectors for each patch to match the notation in \eqref{eq:rot-SC}.

\begin{figure}[!htbp]
  \centering
  \begin{algorithmic}[1]
    \Procedure{R-K-SVD}{$\{\vv{x}_p\}$, $M$, $K$} \State
    $\vv{D} \gets M$ randomly initialized dictionary atoms
 
    \For{fixed number of iterations} \label{alg:main} \State
    $\vv{\mathcal{D}} \gets
    \begin{bmatrix}
      \vv{R}_0 \vv{D} & \vv{R}_1 \vv{D} & \dots & \vv{R}_{R-1} \vv{D}
    \end{bmatrix}$
    \State
    $\{\vv{a}_{p,r}\} \gets \OMP(\{\vv{x}_p\}, \vv{\mathcal{D}}, K)$
    \For{$m = 1, 2, \dots, M$} \State
    $\vv{Y} \gets \begin{bmatrix}\,\end{bmatrix}$
    \For{$p = 1, 2, \dots, P$} \If{$\vv{a}_{p,r}[m] \ne 0$ for some
      $r=r_0$ } \label{ln:r0} \State $\vv{b}_r \gets \vv{a}_{p,r}$ for
    all $r$ \State $\vv{b}_{r_0}[m] \gets 0$ \State
    $\vv{Y} \gets \begin{bmatrix}\vv{Y} & \vv{R}_{-r_0} (\vv{x}_p -
      \vv{\mathcal{D}} \vv{b}) \end{bmatrix}$
    \EndIf
    \EndFor
    \State $\vv{D}_m \gets$ first singular vector of $\vv{Y}$
    \EndFor
    \EndFor
    \State \textbf{return} $\vv{a}$, $\vv{D}$
    \EndProcedure
  \end{algorithmic}
  
  \caption{Pseudocode for the rotational K-SVD algorithm.}
  \label{fig:code}
\end{figure}

The sparse codes having been updated, the second half of the main loop deals with updating the dictionary atoms. 
In order to update atom $m$, all the patches whose sparse code includes a rotated version of $m$ are identified.
These patches are each approximated using the remaining atoms, i.e. $\vv{\mathcal{D}}$ with the column corresponding to the rotated version of $m$ removed.
In normal K-SVD, atom $m$ would be updated to code the residual between these approximations and the original patch;
specifically, atom $m$ is set to the first singular vector of the matrix of residuals.
Instead, we first rotate each residual to an upright orientation relative to atom $m$ before performing the SVD.
That is, if $\vv{R}_{r_0}\vv{D}_m$ codes a patch $\vv{x}_p$, then the patch $\vv{R}_{-r_0}\vv{x}_p$ is upright with respect to $\vv{D}_m$ because it is coded by $\vv{R}_{-r_0}\vv{R}_{r_0}\vv{D}_m=\vv{R}_{0}\vv{D}_m=\vv{D}_m$.
The result of this residual alignment is that the atom is allowed to have a different orientation for each patch it codes,
without these different orientations being averaged during the update step.

It is worth noting a few more implementation details.
First, if the algorithm has not completely converged, it is beneficial to run OMP one more time just before the procedure returns so that the sparse codes correspond to the updated dictionary.
Second, in the dictionary update step, it can happen that a certain atom is used by no patches, meaning that $\vv{Y}$ is empty and the SVD cannot proceed.
This problem is not unique to our algorithm---in fact it happens in K-means clustering as well.
In this case, we simply reset the unused atom to point in the direction of the patch with the highest current coding error.
Third, a single atom may be used to code a given patch more than once (at different orientations).
In this case, a unique $r_0$ cannot be found in line \ref{ln:r0};
in our implementation, we make an arbitrary selection among the possible $r_0$'s.

\subsection{Discrete Steerable Bases}
\label{sec:steerable-basis}
The main challenge of rotational K-SVD is the computational cost.
The OMP step of K-SVD acts on an augmented dictionary that is $R$ times larger than the original dictionary,
and the residuals must be rotated at every iteration.
Because the number of patches to code is typically much larger than the size of the dictionary and because image rotation is more expensive than computing inner products (as occurs inside OMP), it is the rotation of residuals that dominates the cost in practice.
By representing the patches in a discrete steerable basis, we are able to compute the necessary rotations via multiplication with a diagonal steering matrix, thereby accelerating the R-K-SVD.

We motivate our design by considering a unitary $N^2$ by $N^2$ matrix, $\vv{\Phi}$,
that diagonalizes discrete rotations, i.e. $\vv{R}_r = \vv{\Phi} \vv{S}_r \vv{\Phi}^*$,
where $\vv{S}_r$ is a diagonal matrix, which we call the \textit{steering matrix}.
Using $\vv{\Phi}$, which we call a \textit{steerable orthonormal basis},
we could rewrite the error term from \eqref{eq:rot-SC},
\begin{equation}
  \label{eq:rearrange}
  \begin{split}
 \norm*{ \left(\sum_{r=0}^{R-1} \vv{R}_r \vv{D} \vv{a}_{r} \right) - \vv{x} } 
 = \norm*{ \vv{\Phi} \left(\sum_{r=0}^{R-1} \vv{S}_r \vv{\Phi}^* \vv{D} \vv{a}_{r} \right) - \vv{x}} \\
  =  \norm*{ \vv{\Phi} \left(\sum_{r=0}^{R-1} \vv{S}_r  \hat{\vv{D}} \vv{a}_{r} \right) - \vv{\Phi} \hat{\vv{x}}}
  =  \norm*{ \left(\sum_{r=0}^{R-1} \vv{S}_r  \hat{\vv{D}} \vv{a}_{r} \right) -  \hat{\vv{x}}},
\end{split}
\end{equation}
where we use the hat symbol to represent coordinates with respect to $\vv{\Phi}$:
$\vv{\Phi} \hat{\vv{D}} = \vv{D}$ and $\vv{\Phi} \hat{\vv{x}} = \vv{x}$,
and where the last equality uses the fact that $\vv{\Phi}$ is unitary.
This rearrangement is attractive because the rotation matrix $\vv{R}_r$
has been replaced with the diagonal matrix $\vv{S}_r$,
which can be implemented with fewer operations.

In order to design $\vv{\Phi}$, we consider basis functions
\begin{equation}
  \label{eq:basis-cont}
 \phi_{s,t}(r, \theta) = \begin{cases}
e^{\imath t \theta} & \text{if}\quad \frac{N(s-1)}{2S} \le r <  \frac{N s}{2S},\\ 
%
0 & \text{otherwise}.
\end{cases}
\end{equation}
Each basis function, $\phi_{s,t}$, is a circular harmonic function
restricted to be nonzero on a certain annulus;
$s$ is the annulus index and $t$ is the discrete frequency.
These functions are orthogonal and diagonalize rotations,
\begin{equation}
  \label{eq:rotation}
\phi_{s,t}(r, \theta - \theta_0) = e^{-\imath  t \theta_0} \phi_{s,t}(r, \theta).
\end{equation}

\begin{figure}[!htbp]
  \centering
  \begin{minipage}[b]{.5\linewidth}
     \subfloat[$s=1$]{\includegraphics[width=.31666\figwidth]{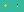}}\\
     \subfloat[$s=2$]{\includegraphics[width=.475\figwidth]{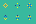}}
   \end{minipage}\hfill
   \subfloat[$s=3$]{\includegraphics[width=.475\figwidth]{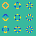}}
   \caption{Example discrete steerable basis with $S=3$ and $T_1=1$, $T_2=5$, and $T_3=8$.
   Plotted here are the real parts of basis elements with non-negative discrete frequency.}
  \label{fig:basis}
\end{figure}

To form the discrete steerable basis,
we sample these functions at grid locations $\vv{k} \in \{-(N-1)/2, \dots, (N-1)/2\} \times  \{-(N-1)/2, \dots, (N-1)/2\}$
such that, $\vv{\phi}_{s,t}[\vv{k}] = \phi_{s,t}(\norm{\vv{k}}, \angle \vv{k})$
(note that $\vv{\phi}_{s,t}$ on the left hand side is bold,
indicating that it is a vector).
We normalize the resulting vectors,
and stack them column-wise,
\begin{equation}
  \label{eq:steerable-basis}
 \vv{\Phi} =
\begin{bmatrix}
  \vv{\phi}_{1,-T_1} & \dots & \vv{\phi}_{1, T_1} &  \dots & \vv{\phi}_{S, -T_S} & \dots & \vv{\phi}_{S,T_S}\end{bmatrix};
\end{equation}
see Figure~\ref{fig:basis} for an example.
The steering matrix, $\vv{S}_r$, then has diagonal elements which we can read off from \eqref{eq:rotation},
\begin{multline}
  \label{eq:Sr}
  \left[\begin{matrix}
      e^{-\imath (-T_1) \frac{2\pi r}{R}}& \dots & e^{-\imath T_1 \frac{2\pi r
          }{R}}  & e^{-\imath (-T_2) \frac{2\pi r}{R}} & \dots
    \end{matrix} \right.\\
  \left. \begin{matrix}
      e^{-\imath T_2 \frac{2\pi r }{R}} & \dots & e^{-\imath (-T_S)
        \frac{2\pi r}{R}} & \dots& e^{-\imath  T_S\frac{2\pi r}{R}} 
    \end{matrix} \right].
\end{multline}

There are several problems that arise when performing this sampling.
First, the maximum discrete frequencies, $\{T_s\}$, must be kept low enough to avoid aliasing.
We find that setting $T_s$ equal to half the circumference of the $s$-th annulus (measured in pixels) is sufficient;
this choice is consistent with the Nyquist criterion.
Similarly, $S$ must not be too large, we use $\lfloor \frac{N}{2} \rfloor$.
With such settings for $S$ and $\{T_s\}$,
the number of columns in $\vv{\Phi}$ is generally lower than $N^2$;
this means that $\vv{\Phi}$ spans a subspace of $\mathbb{C}^{N^2}$.
At the same time, even with conservative settings for $S$ and $\{T_s\}$,
$\vv{\Phi}$ cannot be exactly orthonormal when discretized.
Instead, $\vv{\Phi}$ and $\tilde{\vv{\Phi}} = (\vv{\Phi}^*\vv{\Phi})^{-1} \vv{\Phi}^*$ form a \textit{steerable biorthogonal pair of bases}.
But, because $\vv{\Phi}$ is nearly orthogonal,
we know that there exist constants such that
$\lambda_{\text{min}} \norm{\vv{x}} \le \norm{\tilde{\vv{\Phi}} \vv{x}} \le \lambda_{\text{max}} \norm{\vv{x}}$ with
$\lambda_{\text{min}} \approx \lambda_{\text{max}} \approx 1$.

\subsection{Proposed Algorithm}
In summary, our proposed algorithm is as follows.
\begin{inparaenum}
\item Generate the steerable basis, $\vv{\Phi}$, according to \eqref{eq:basis-cont} and \eqref{eq:steerable-basis}.
\item Compute coordinates of the input patches with respect to $\vv{\Phi}$,  $\hat{\vv{x}}_p=(\vv{\Phi}^* \vv{\Phi})^{-1} \vv{\Phi}^* \vv{x}_p$.
\item Run the rotational K-SVD algorithm on $\{\hat{\vv{x}}_p\}$,
  replacing all instances of the matrix $\vv{R}_r$ with the diagonal matrix $\vv{S}_r$, with elements given in \eqref{eq:Sr}.
  The R-K-SVD will return $\hat{\vv{D}}$, the coordinates of $\vv{D}$ with respect to $\vv{\Phi}$.
  \item  Return the learned dictionary to the patch domain via $\vv{D} = \vv{\Phi} \hat{\vv{D}}$.
\end{inparaenum}

The advantage of this approach over R-K-SVD without the coordinate transform is, again,
that $\vv{S}_r$ is diagonal, while $\vv{R}_r$ is not (though it may be sparse, depending on implementation).
The only drawback is that a small approximation is made when rearranging the data term~\eqref{eq:rearrange}
because of the non-orthogonality of $\vv{\Phi}$.

\section{Experiments and Results}
\label{sec:experiments}
We now present experiments and results on sparse patch coding, texture classification, and image rotation.

\subsection{Sparse Patch Coding}\label{sec:imageCoding}
We compare the image patch coding performance of the standard formulation~(\ref{eq:SC}) and the proposed rotational formulation~(\ref{eq:rot-SC}).
We perform the standard sparse coding using the \texttt{mexTrainDL} function from the sparse modeling software (SPAMS)~\cite{MBP2014}%
\footnote{\texttt{http://spams-devel.gforge.inria.fr}, as of April 2018.},
using OMP (i.e., \texttt{param.mode=1}), a batch size of 400 patches and 1000 iterations.
We perform rotational sparse coding using ten iterations of R-K-SVD (Figure~\ref{fig:code}) with 60 steering angles per atom.

Circular patches are considered.
As an example, a patch diameter of 11 includes a total of 97 pixels.
The corresponding number of elements of the steerable basis $\vv{\Phi}$ is 95.
The mean squared error (MSE) is used to quantify the coding performance.

{\bf Synthetic Images.}
In a first experiment, we evaluate rotation invariance and subsequent representational power of the proposed coding approach.
To this end, we generated a set of 1000 patches obtained from the sum of two random rotations of a single ``true'' atom (see Figure~\ref{fig:toyCoding} (a) and (b)).
A patch diameter of 42 is used.
For standard sparse coding, we try $M=1, K=1$ and $M=2, K=2$.
For rotational sparse coding, we set $M=1, K=2$.
Whereas standard sparse coding yields dictionary atoms that are angular averages of the patches (Figure~\ref{fig:toyCoding} (c) and (e)),
rotational sparse coding correctly identifies the underlying template (Figure~\ref{fig:toyCoding}(g)).
This results in increased coding performance even when using one single steerable atom (\mbox{MSE = 3.64}),
where using two atoms of K-SVD yields \mbox{MSE = 10.35}.

\begin{figure}[!htbp]
\centering
\subfloat[\textit{cameraman} (256 by 256)]{\includegraphics[trim = 0 0 0 0, clip, width=.45\figwidth]{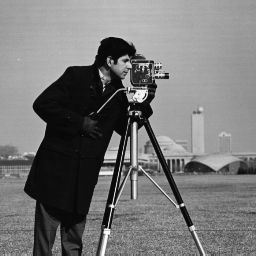}}\hfill
\subfloat[\textit{net} (640 by 640)]{\includegraphics[trim = 0 0 0 0, clip, width=.45\figwidth]{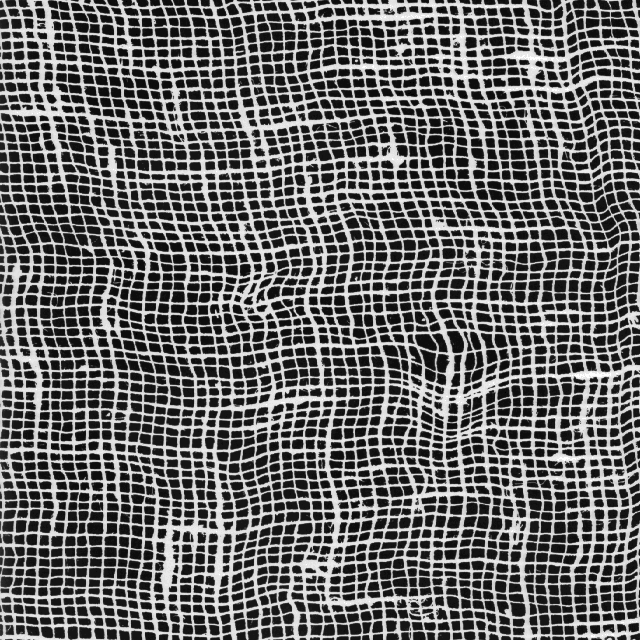}}
  \caption{Images used to compare the coding performance of R-K-SVD and the standard K-SVD in Figures~\ref{fig:MSEcameraman} and~\ref{fig:MSEnet}.}
  \label{fig:MSEimages}
\end{figure}

{\bf Natural Images.}
Second, we investigate the influence of the number of atoms $M$ and sparsity $K$ for coding image patches from two natural images,
\textit{cameraman} and \textit{net}\footnote{The image is taken from the Brodatz collection~\cite{Bro1966}, referred to as ``D103''.}
(Figure~\ref{fig:MSEimages}).
Both images have pixel values in $[0,1]$.
Sliding patches with a stride of 1 and a diameter of 11 pixels are used.
Pixel values are rescaled to obtain a unit norm at the patch level.
The R-K-SVD lower bound (LB) is computed as the error obtained from the direct projection of the patches on the discrete steerable basis
and illustrates the span of the latter.

\begin{figure}[!htbp]
\centering
\subfloat[coding error]{\includegraphics[trim = 80 270 95 285, clip, width=\figwidth]{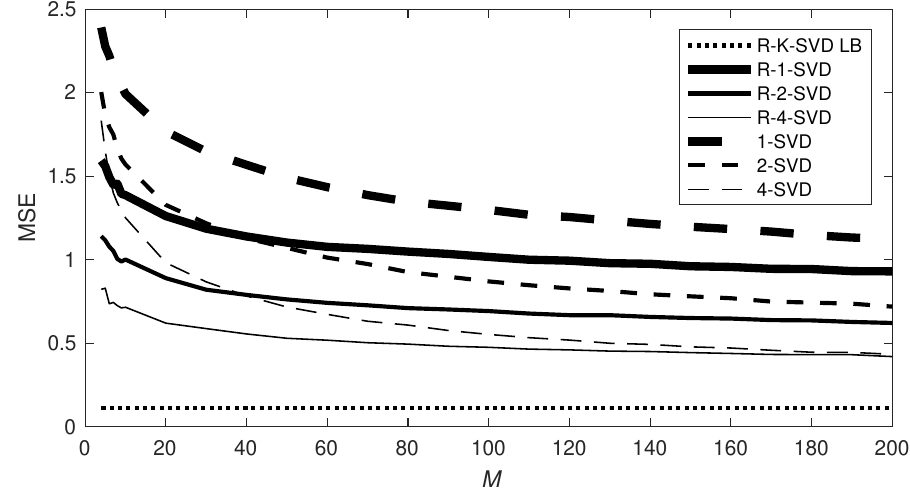}}\\
\subfloat[image patches]{%
   \begin{tikzpicture}[color1/.style={red,ultra thick,rounded corners,opacity=0.5}]
      \node[anchor=south west,inner sep=0] (image) at (0,0) {\includegraphics[width=\figwidth]{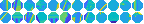}};
      \begin{scope}[x={(image.south east)},y={(image.north west)}]
        \draw[color1] (0 * 1/12 + 1/12/2 -.003, 0 *1/2+1/2/2 +.012) circle (.36cm);
        \draw[color1] (1 * 1/12 + 1/12/2 -.003, 1 *1/2+1/2/2 +.012) circle (.36cm);
        \draw[color1] (2 * 1/12 + 1/12/2 -.003, 0 *1/2+1/2/2 +.012) circle (.36cm);
        \draw[color1] (8 * 1/12 + 1/12/2 -.003, 1 *1/2+1/2/2 +.012) circle (.36cm);
    \end{scope}
    \end{tikzpicture}}\\
\subfloat[K-SVD coding, MSE 0.98]{ \begin{tikzpicture}[color1/.style={red,ultra thick,rounded corners,opacity=0.5}]
      \node[anchor=south west,inner sep=0] (image) at (0,0) {\includegraphics[width=\figwidth]{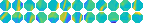}};
      \begin{scope}[x={(image.south east)},y={(image.north west)}]
        \draw[color1] (0 * 1/12 + 1/12/2 -.003, 0 *1/2+1/2/2 +.012) circle (.36cm);
        \draw[color1] (1 * 1/12 + 1/12/2 -.003, 1 *1/2+1/2/2 +.012) circle (.36cm);
        \draw[color1] (2 * 1/12 + 1/12/2 -.003, 0 *1/2+1/2/2 +.012) circle (.36cm);
        \draw[color1] (8 * 1/12 + 1/12/2 -.003, 1 *1/2+1/2/2 +.012) circle (.36cm);
    \end{scope}
    \end{tikzpicture}}\\
\subfloat[R-K-SVD coding, MSE 0.61]{ \begin{tikzpicture}[color1/.style={red,ultra thick,rounded corners,opacity=0.5}]
      \node[anchor=south west,inner sep=0] (image) at (0,0) {\includegraphics[width=\figwidth]{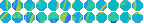}};
      \begin{scope}[x={(image.south east)},y={(image.north west)}]
        \draw[color1] (0 * 1/12 + 1/12/2 -.003, 0 *1/2+1/2/2 +.012) circle (.36cm);
        \draw[color1] (1 * 1/12 + 1/12/2 -.003, 1 *1/2+1/2/2 +.012) circle (.36cm);
        \draw[color1] (2 * 1/12 + 1/12/2 -.003, 0 *1/2+1/2/2 +.012) circle (.36cm);
        \draw[color1] (8 * 1/12 + 1/12/2 -.003, 1 *1/2+1/2/2 +.012) circle (.36cm);
    \end{scope}
    \end{tikzpicture}}\\
  \begin{minipage}[b]{1\linewidth}
  \centering
  \vspace{.25cm}
  \subfloat[standard atoms]{%
    \begin{tikzpicture}[color1/.style={blue,ultra thick,rounded corners,opacity=0.5}]
      \node[anchor=south west,inner sep=0] (image) at (0,0) {\includegraphics[trim = 0 0 0 0, clip, width=.41666\figwidth]{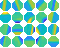}};
      \begin{scope}[x={(image.south east)},y={(image.north west)}]
        \draw[color1] (0 *.2+.1-.01, 0 *.25+.125 + .01) circle (.36cm);
        \draw[color1] (1 *.2+.1-.01, 0 *.25+.125 + .01) circle (.36cm);
        \draw[color1] (0 *.2+.1-.01, 1 *.25+.125 + .01) circle (.36cm);
        \draw[color1] (2 *.2+.1-.01, 0 *.25+.125 + .01) circle (.36cm);
    \end{scope}
    \end{tikzpicture}}
     \hspace{1cm}
     \subfloat[steerable atoms]{\includegraphics[width=.41666\figwidth]{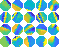}}
 \end{minipage}
  \caption{Coding performance for \textit{cameraman} based on 17,330 sliding patches. 
  Quantitative comparison of the coding error (a) for K-SVD and R-K-SVD. 
  A qualitative comparison of coding (b-d) and atoms (e,f) is shown for \mbox{$M=20, K=4$}.
  Red outlines highlight patches where the advantages of R-K-SVD are clearest.
  Blue outlines show atoms in the standard dictionary that are related by rotation.
}
  \label{fig:MSEcameraman}
\end{figure}
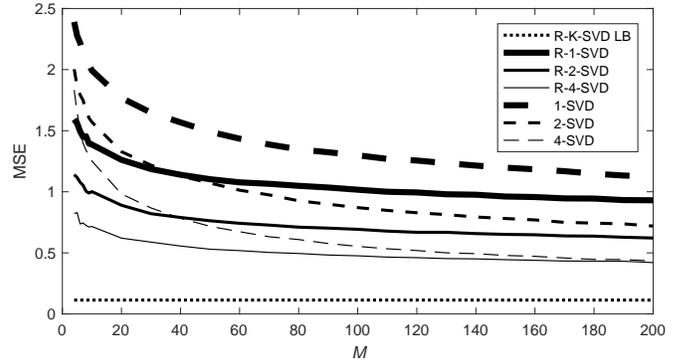
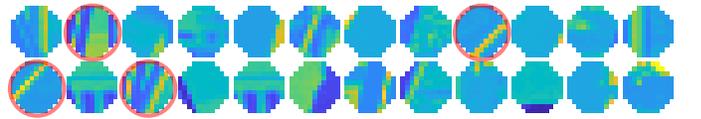
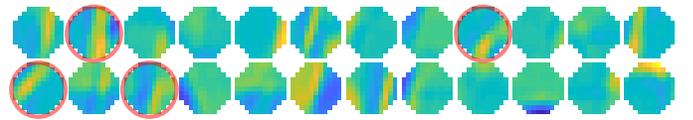
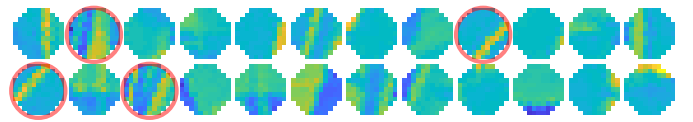
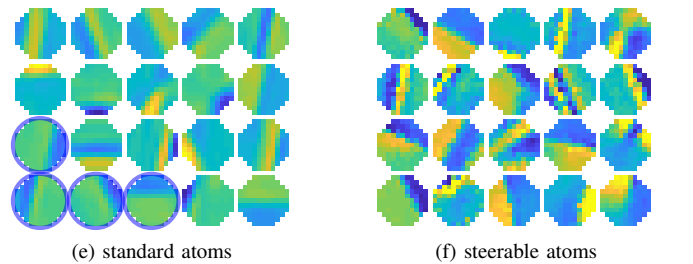

\begin{figure}[!htbp]
\centering
\subfloat[coding error]{\includegraphics[trim = 80 270 95 285, clip, width=\figwidth]{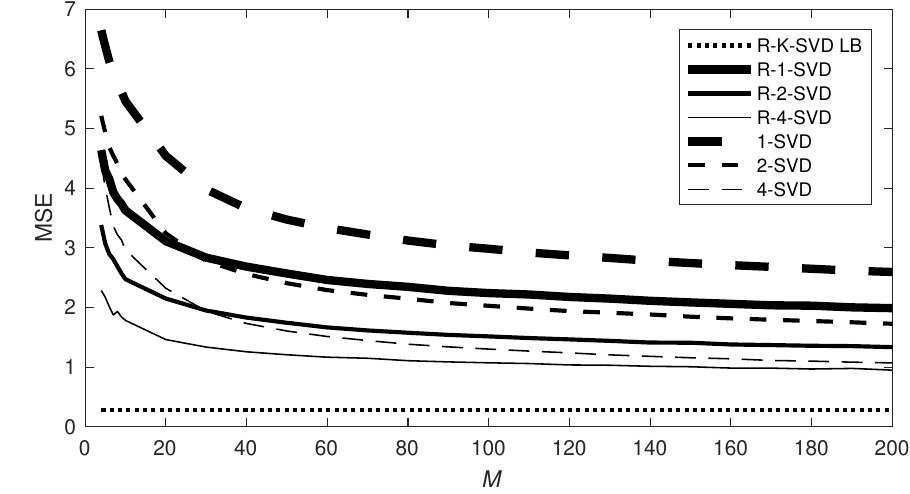}}\\
\subfloat[image patches]{\includegraphics[width=\figwidth]{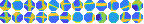}}\\
\subfloat[K-SVD coding, MSE 4.13]{\includegraphics[width=\figwidth]{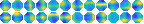}}\\
\subfloat[R-K-SVD coding, MSE 2.51]{\includegraphics[width=\figwidth]{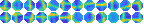}}\\
  \begin{minipage}[b]{1\linewidth}
  \centering
  \vspace{.25cm}
     \subfloat[standard atoms]{\includegraphics[trim = 0 0 0 0, clip, width=.41666\figwidth]{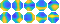}}
     \hspace{1cm}
     \subfloat[steerable atoms]{\includegraphics[width=.41666\figwidth]{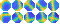}}
 \end{minipage}
  \caption{Coding performance for \textit{net} based on 396,855 sliding patches.
  Quantitative comparison of the coding error (a) for K-SVD and R-K-SVD. 
  A qualitative comparison of coding (b-d) and atoms (e,f) is shown for \mbox{$M=10, K=2$}.}
  \label{fig:MSEnet}
\end{figure}

The results (Figures \ref{fig:MSEcameraman} and \ref{fig:MSEnet}) show that the coding error for the rotational formulation is consistently lower
than for the standard formulation, though the gap between them decreases as the dictionary size increases.
It is worth noting that, for fixed $K$ and $M$, the rotational formulation requires more bits to code a patch
than the standard one, because it must store $K$ atom indices and $K$ rotation indices.
But, even accounting for this by comparing, e.g., standard coding with $K=2$ to rotational coding with $K=1$,
we see that rotational coding still produces lower error under $M \approx 30$.
Qualitatively, the rotational codings and atoms are much sharper than the standard ones.
We also see that some of the standard atoms are nearly rotations of each other,
which does not happen in the rotational case.

\begin{figure}[!htbp]
\centering
\includegraphics[trim = 80 295 95 305, clip, scale=0.57]{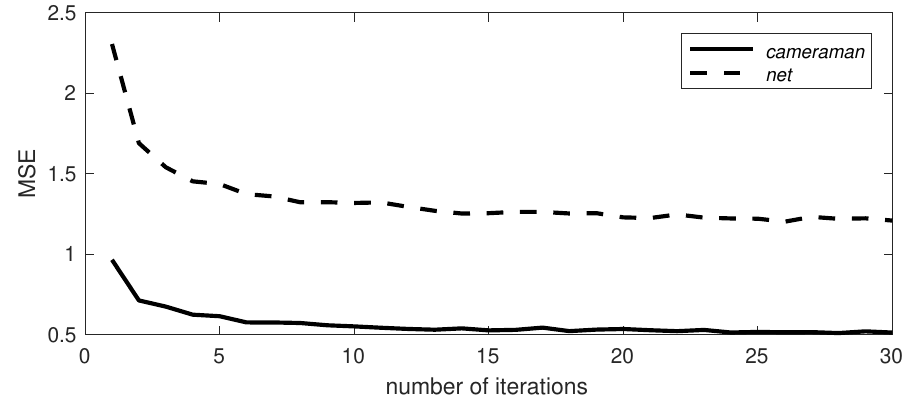}\\
\caption{Convergence with respect to the number of iterations (see Figure~\ref{fig:code}) for \mbox{$M=20, K=5$}.}
  \label{fig:MSE_iters}
\end{figure}

\begin{figure}[!htbp]
\centering
\includegraphics[trim = 80 295 95 305, clip, scale=0.57]{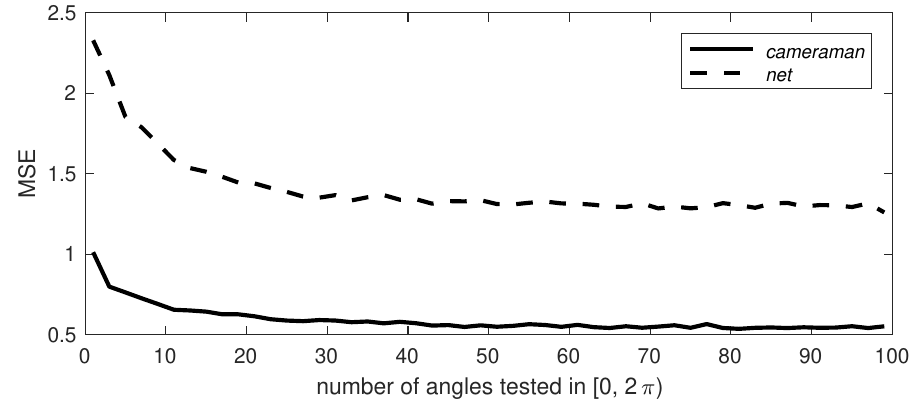}\\
\caption{Influence of the number of angles tested in $[0,2\pi)$ for \mbox{$M=20, K=5$}.}
  \label{fig:MSE_thetas}
\end{figure}

{\bf Parameter Sensitivity.}
In a third coding experiment,
we study parameter sensitivity with regard to the number of iterations of R-K-SVD
and number of angles tested in $[0,2\pi)$ in Fig.~\ref{fig:MSE_iters} and \ref{fig:MSE_thetas}, respectively.
For both \textit{cameraman} and \textit{net}, the optimal performance is reached after 10 iterations and 40 angles tested.

\subsection{Texture Classification}
\label{sec:texture}
\begin{figure}[!htbp]
\includegraphics[width=\figwidth/8,trim=0 0 1px 1px,clip]{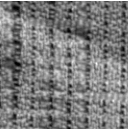}%
\includegraphics[width=\figwidth/8,trim=0 0 1px 1px,clip]{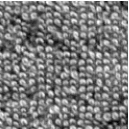}%
\includegraphics[width=\figwidth/8,trim=0 0 1px 1px,clip]{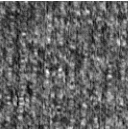}%
\includegraphics[width=\figwidth/8,trim=0 0 1px 1px,clip]{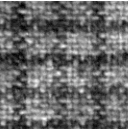}%
\includegraphics[width=\figwidth/8,trim=0 0 1px 1px,clip]{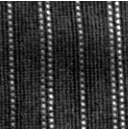}%
\includegraphics[width=\figwidth/8,trim=0 0 1px 1px,clip]{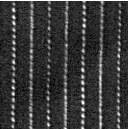}%
\includegraphics[width=\figwidth/8,trim=0 0 1px 1px,clip]{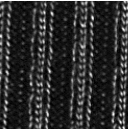}%
\includegraphics[width=\figwidth/8,trim=0 0 1px 1px,clip]{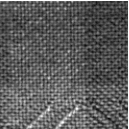}\\%
\includegraphics[width=\figwidth/8,trim=0 0 1px 1px,clip]{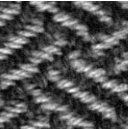}%
\includegraphics[width=\figwidth/8,trim=0 0 1px 1px,clip]{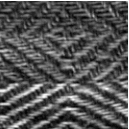}%
\includegraphics[width=\figwidth/8,trim=0 0 1px 1px,clip]{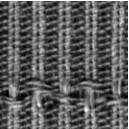}%
\includegraphics[width=\figwidth/8,trim=0 0 1px 1px,clip]{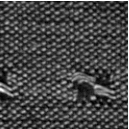}%
\includegraphics[width=\figwidth/8,trim=0 0 1px 1px,clip]{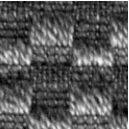}%
\includegraphics[width=\figwidth/8,trim=0 0 1px 1px,clip]{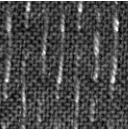}%
\includegraphics[width=\figwidth/8,trim=0 0 1px 1px,clip]{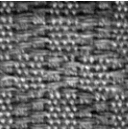}%
\includegraphics[width=\figwidth/8,trim=0 0 1px 1px,clip]{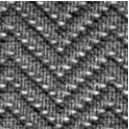}\\%
\includegraphics[width=\figwidth/8,trim=0 0 1px 1px,clip]{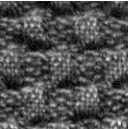}%
\includegraphics[width=\figwidth/8,trim=0 0 1px 1px,clip]{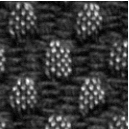}%
\includegraphics[width=\figwidth/8,trim=0 0 1px 1px,clip]{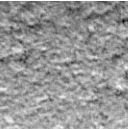}%
\includegraphics[width=\figwidth/8,trim=0 0 1px 1px,clip]{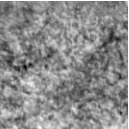}%
\includegraphics[width=\figwidth/8,trim=0 0 1px 1px,clip]{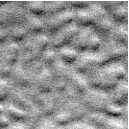}%
\includegraphics[width=\figwidth/8,trim=0 0 1px 1px,clip]{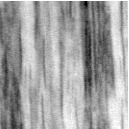}%
\includegraphics[width=\figwidth/8,trim=0 0 1px 1px,clip]{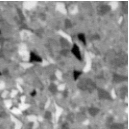}%
\includegraphics[width=\figwidth/8,trim=0 0 1px 1px,clip]{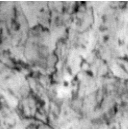}%
\caption{Upright samples of the 24 texture classes included the Outex\_TC\_10 validation suite.
  Each image is 128 by 128.}\label{fig:outexDB}
\end{figure}

We also explore the advantage of using rotational sparse coding in a texture classification application.
Speaking generally, the task is to classify an input image as belonging to one of several texture classes (e.g., \textit{sand}, \textit{pebbles}),
assuming access to a training set with a number of exemplar images from each class.
See Figure~\ref{fig:outexDB}
for examples of texture classes.


Texture classification is amenable to dictionary learning-based approaches
because textures often contain repeated elements
(called microtextures or textons)
that can be learned.
Thus, a dictionary learned on examples of a given class should contain some discriminative information about that class.
A set of dictionaries, one learned on each class, should be useful for classification.
Our hypothesis is that,
in this context,
the rotational dictionary should be superior to the standard one because it is able to achieve small coding errors with fewer atoms,
meaning that the per-class dictionaries can be very discriminative.
Stated another way,
we hypothesize that textons repeat not only at different locations,
but also with different rotations,
and therefore rotational dictionary learning will be
able to represent them more easily.

{\bf Dataset.}
We evaluated the texture classification performance on the Outex Database~\cite{OMP2002}.
Outex
 contains photographs of materials (e.g., \textit{fabric}, \textit{tile}) acquired with a standardized protocol.
In particular, the Outex\_TC\_10 validation suite is the most commonly used in the literature and contains 24 texture classes (see Figure~\ref{fig:outexDB}).
Considered texture samples are 128 by 128-pixel images cropped from physically translated and rotated versions of larger samples.
The training set contains 480 images including 20 upright samples per class.
The test set includes $24$ rotated samples where angles considered are ($5^{\circ}$, $10^{\circ}$, $15^{\circ}$, $30^{\circ}$, $45^{\circ}$, $60^{\circ}$, $75^{\circ}$, and $90^{\circ}$) and where each angle index contains 20 samples per class, resulting in 3840 test images.
We expect this to be a good fit for rotational dictionary learning because
the construction of the testing set explicitly rewards rotation-invariant methods.

%

{\bf Classifier.}
Our classification approach is as follows.
First, we learn one sparse dictionary per class from the training images.
To do this, we split each training image into sliding patches (stride of one pixel).
For each class, we randomly select 4,000 of patches from each associated training image
resulting in a dataset of 80,000 patches for sparse coding 
After thus learning one dictionary for each class,
we use these dictionaries to code every patch from each training image.
For each patch, we assign the class label corresponding to the class dictionary
that codes the patch with the minimum sum squared error.
We aggregate these patch-level classifications into a histogram for each training image;
this histogram is a 24 by 1 feature vector for each training image.
To classify a testing image, we form a histogram in the same way
and classify the image based on its nearest neighbor in the training set
(using the $\chi^2$ distance: $d(a,b) = \sum_{i} (a[i] - b[i])^2/(a[i] + b[i])$). 

We compare three variations of this approach:
one using standard sparse coding (K-SVD),
one using standard sparse coding with an augmented training set (aug-K-SVD),
and one using rotational sparse coding (R-K-SVD).
For the augmented approach,
during dictionary learning, each training patch takes a random rotation in the set \{$5^{\circ}$, $10^{\circ}$, $15^{\circ}$, $30^{\circ}$, $45^{\circ}$, $60^{\circ}$, $75^{\circ}$, $90^{\circ}$\},
which mimics the test data.
This is a common and effective strategy for adding invariance to classifiers
(e.g., artificial neural networks);
it essentially teaches the classifier that rotations should be ignored.

For all three approaches,
the parameters to set are the size of the dictionary, $M$,
the  patch size, $N$, and the sparsity level, $K$.
These are jointly swept over the values  $\{10, 25, 50, 100\}$, $\{11, 13, 15\}$, and  $\{1, 2, 3\}$, respectively.
From these 36 parameter settings, the best combination is selected via cross-validation on the training set:
after dictionary learning, a random subset comprising 88\% of the training images is held out and classified
(the 88\% ratio is selected to because it is the ratio of training to test data).
This procedure is repeated 100 times and the accuracies of each fold averaged.
The parameter setting with the highest average cross-validation accuracy is used during testing.
The selected parameters for 
K-SVD were M=100, N=11, K=3;
for augmented K-SVD were M=100, N=15, K=1;
and for R-K-SVD were M=100, N=15, K=1.

{\bf Comparison methods.}
While there are numerous published results on Outex
(see, e.g., \cite{depeursinge_steerable_2017}
for a table with over a dozen methods from the last two decades),
we also ran our own comparisons.
Specifically, we compared our approach with current state of the art deep learning frameworks trained end-to-end.
We compared seven architectures with and without pretraining and rotational data augmentation
(for more details, see Appendix~\ref{app:deep_results}).
The best results, obtained with ResNet50~\cite{he2016deep}, InceptionV3~\cite{szegedy2016rethinking} and DenseNet169~\cite{HLW2017} are reported as comparisons in Table~\ref{tab:tex-results}.
As an additional comparison, we used the approach proposed by Cimpoi \textit{et al}.~\cite{Cimpoi2016} based on Fisher Vectors and CNNs (FV-CNN).
After resampling the images to 231 by 231 using bicubic interpolation, OverFeat~\cite{Sermanet2013} was used to extract deep 1024 feature maps from AlexNet (\textit{fast} model) at the last convolutional layer.
FVs were then computed following the setup proposed in~\cite{Cimpoi2016} with 64 Gaussian mixture components using VLfeat~\cite{vedaldi08vlfeat}.
We retained the first 64 principal components,
resulting in a length-64 feature vector for each image.
(This length was chosen to maximize testing accuracy;
the values we tried were 24, 32, 64, 128, and 256.)
We classify testing images by their nearest neighbor
in the training set
(here using Euclidean distance because the feature vectors are not histograms).
We repeat the same procedure for the augmented training set
as described in the previous subsection.

{\bf Results.}
We report the results in terms of testing accuracy in Table~\ref{tab:tex-results}.
For additional context, we include six recently-reported results%
---%
again,
see \cite{depeursinge_steerable_2017} for a more extensive table.
The accuracy of our rotational sparse coding-based classifier is 100\%,
which we believe is the first time a perfect result has been reported
for the Outex\_TC\_10 dataset.
The next best result in the literature comes from an artificial neural network-based approach \cite{marcos_learning_2016},
which reports 99.95\% (a misclassification of only two testing images),
and our implementation of DenseNet169 achieved 99.98 (missing only one testing image).
So, while the present results are not a tremendous improvement in accuracy,
it is remarkable that our relatively simple approach achieves perfect performance,
especially given the number of methods that have been tried.
We also show that standard dictionary learning performs significantly better than chance (1/24 = 0.0417),
and that augmenting the training set does improve the accuracy.
Results for the FV-CNN are significantly better than for the standard dictionary learning,
but they do not approach the state of the art,
even with data augmentation.
This may indicate that, while the FV-CNN captures useful image features,
data augmentation alone is not sufficient to provide the rotation invariance needed for Outex.
These results indicate that the rotational formulation of dictionary learning
is useful in classifying rotation-invariant textures.

\begin{table}[!htbp]
  \centering
    \begin{threeparttable}
  \caption{Texture classification accuracy on Outex\_TC\_10}
  \begin{tabular}{rr}\toprule
    method &  accuracy\\ \midrule
    Hadizadeh                 2015~\cite{Had2015}                    &  0.9730  \\ 
    Shrivastava \textit{et al.} 2015~\cite{ShT2015}                    &  0.9919 \\
    Zand \textit{et al.}        2015~\cite{ZDH2015}                    &  0.9838 \\
    Liu \textit{et al.}         2016~\cite{LLF2016}                    & 0.9987  \\
    Marcos \textit{et al.}      2016 \cite{marcos_learning_2016}       & 0.9995   \\
    Depeursinge \textit{et al.} 2017~\cite{depeursinge_steerable_2017} & 0.9956   \\\midrule
    K-SVD     &     0.4432                                           \\
    aug-K-SVD  &     0.5635                                          \\
    FV-CNN (64 features)~\cite{Cimpoi2016}     &     0.5471 \\
    FV-CNN-aug (64 features)~\cite{Cimpoi2016} &     0.6487  \\
    ResNet50~\cite{he2016deep} &     0.9945 \\
    InceptionV3~\cite{szegedy2016rethinking} &     0.9944 \\
    DenseNet169~\cite{HLW2017} &     0.9983 \\
    R-K-SVD   & {\bf 1.0000} \\
\bottomrule
  \end{tabular}
    \begin{tablenotes}
    \item {The top seven rows are results reported in other papers.
        The bottom eight rows are results from experiments in this paper,
        with our own implementations of the corresponding methods.}
    \end{tablenotes}
  \end{threeparttable}
\label{tab:tex-results}
\end{table}

To shed more light on the sparse coding methods,
we comment here on the results of the parameter sweeps
described above%
---%
see Table~\ref{tab:tex-param} in Appendix~\ref{app:sweep} for the full results.
First,
the cross-validation accuracy is highest for standard sparse coding,
while its  testing performance is the lowest.
This is consistent with overfitting:
because training and validation both occur only on upright images,
the standard dictionary fits upright appearance;
the other two methods specifically learn in a rotational way,
which naturally sacrifices some training performance.
Second,
the augmented and rotational methods generally benefit from using large patches ($N=15$),
while there is no clear trend for the standard method.
The underlying tradeoff is that, when larger patches are used,
there is effectively less training data (because an image contains fewer independent large patches)
and it is higher dimensional (because there are more pixels in each patch).
Larger patches also slow down the training and testing stages.
Third,
all the methods have the highest cross-validation accuracy for the largest dictionary, $M=100$.
The dictionary size, $M$, and the sparsity, $K$,
control the balance between the representation and discrimination abilities of the dictionary.
Larger $M$ and $K$ mean more representation power,
which decreases the coding error for patches of the chosen class.
But, larger $M$ and $K$ also reduce the coding error for patches outside the chosen class,
reducing discrimination.
The results suggest that single textures are heterogeneous enough
to require a large dictionary to describe them well.


\subsection{Rotation Speed}
\begin{figure}
  \centering
  \subfloat[original]{\includegraphics[width=.2\figwidth]{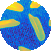}}%
  \subfloat[nearest]{\includegraphics[width=.2\figwidth]{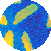}}%
  \subfloat[bicubic]{\includegraphics[width=.2\figwidth]{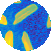}}%
  \subfloat[steering]{\includegraphics[width=.2\figwidth]{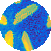}}%
  \subfloat[steering$_{\text{low}}$]{\includegraphics[width=.2\figwidth]{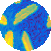}}%
  \caption{Comparison of quality between nearest neighbor, bicubic, and the proposed steering-based rotation.
  Steering is comparable to bicubic interpolation, while the nearest neighbor interpolation exhibits staircase artifacts.}
  \label{fig:ex-rotation}
\end{figure}

In our final experiment,
we compare the proposed steering-based patch rotation
to standard rotations using interpolation.
For a fixed patch size, $N$ by $N$,
we form matrices $\vv{R}_{\text{nearest}}$ and $\vv{R}_{\text{bicubic}}$,
which are $N^2$ by $N^2$ matrices that perform a rotation by $100 + \sqrt{2}$ degrees
using nearest neighbor and bicubic interpolation, respectively.
We also create a steerable basis, $\vv{\Phi}$ and steering matrix $\vv{S}$,
as described in Section~\ref{sec:steerable-basis},
which achieves the same rotation.
Finally, we create a lowpass steerable basis, $\vv{\Phi}_{\text{low}}$, and steering matrix $\vv{S}_{\text{low}}$,
where the cutoff frequencies $T_s$ have been reduced by half as compared to $\vv{\Phi}$.
Thus, for a vectorized patch, $\vv{x}$, rotation can be performed by one of
$\vv{R}_{\text{nearest}} \vv{x}$,
$\vv{R}_{\text{bicubic}} \vv{x}$,
$\vv{\Phi} \vv{S} (\vv{\Phi}^* \vv{\Phi})^{-1} \vv{\Phi}^* \vv{x}$,
or  $\vv{\Phi}_{\text{low}} \vv{S}_{\text{low}} (\vv{\Phi}_{\text{low}}^* \vv{\Phi}_{\text{low}})^{-1} \vv{\Phi}_{\text{low}}^* \vv{x}$.
We compare these four types of rotation on a 51 by 51 patch in Figure~\ref{fig:ex-rotation}.
The steering-based rotation is qualitatively similar to bicubic interpolation,
and it avoids the staircase artifacts present in the nearest neighbor approach.
Reducing the cutoff frequencies gives a smoother result.

\begin{figure}
  \centering
  \includegraphics[width=229.8775pt]{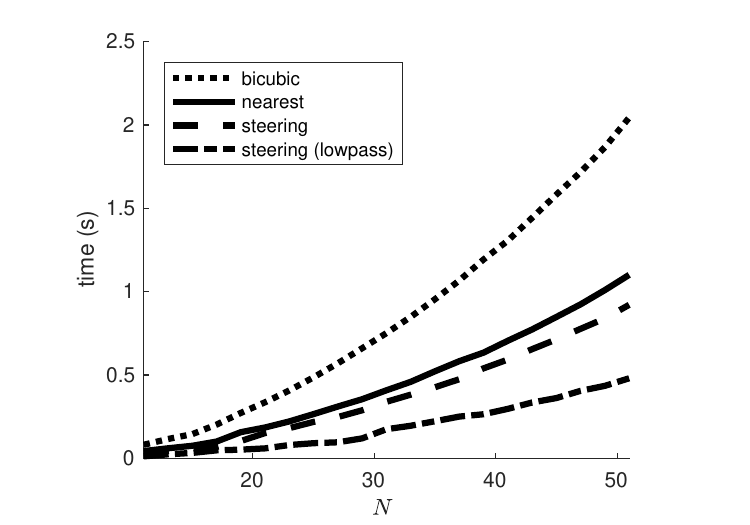}
  \caption{Timing results for rotating 50,000 patches of size $N$ by $N$.
  Rotation using steerability is significantly faster than bicubic interpolation.}
  \label{fig:results-rotation}
\end{figure}

In order to compare these three rotation methods in the context of the R-K-SVD algorithm,
we generate a set of 50,000 random patches and concatenate them into an $N^2$ by 50,000 matrix, $\vv{X}$.
We then measure the time it takes to left multiply this matrix by $\vv{R}_{\text{nearest}}$ and $\vv{R}_{\text{bicubic}}$.
We also compute $\hat{\vv{X}} =  (\vv{\Phi}^* \vv{\Phi})^{-1} \vv{\Phi}^* \vv{X}$
and measure the time it takes to left multiply this by $\vv{S}$ (and likewise for $\vv{\Phi}_{\text{low}}$ and  $\vv{S}_{\text{low}}$).
The rationale for this comparison is that finding $\hat{\vv{X}}$ only occurs once in our algorithm,
while the multiplication by $\vv{R}$ or $\vv{S}$ must happen every iteration.
All matrix multiplications are done using \texttt{sparse} matrices in Matlab,
and timing is performed with the \texttt{timeit} function.
The results (Figure~\ref{fig:results-rotation}) show that using steering for rotation is faster than either
nearest neighbor or bicubic interpolation,
with steering being consistently more than twice as fast as bicubic interpolation.

\section{Discussions and Conclusion}
\label{sec:discussion}
We presented a novel, fast rotational sparse coding approach based on steerability.
Our experiments showed that the rotational formulation provides advantages over
the standard sparse coding formulation for coding and texture classification.
In particular, its ability to exploit the rotational symmetry of image patches
allows the span of the learned dictionary to be reduced
without compromising its representational power.

Much of image processing involves either cleverly exploiting
or laboring to handle various invariances and equivariances.
An excellent example of this theme is convolutional neural networks (CNNs),
which, by relying on convolution, both provide useful invariance to translations
as well as allow for smaller and faster models than fully-connected networks.
For this reason, we believe that the ideas presented here on rotational sparse coding may be useful
elsewhere in image processing, especially because the steerable basis can provide
a very efficient implementation of rotation in many contexts beyond sparse coding.
Indeed, there is already some work on adding steerability to CNNs~\cite{weiler_learning_2018,andrearczyk2019exploring}.

The design of discrete steerable bases remains an open topic.
It is not clear whether a steerable orthonormal basis can be found,
but it is likely that the ad hoc design described in Section~\ref{sec:steerable-basis} can be improved upon.
And, even using our specific design, there are several parameters
that could be explored further experimentally.
We also note that reducing the maximum frequencies provides
an easy way to trade rotation quality for speed.

The coding experiments give a sense of how much more expressive
rotational sparse coding is than the standard formulation.
In particular, we observed for both \textit{cameraman} and \textit{net} images that R-K-SVD provides a lower MSE than standard K-SVD, even when the former uses $K=1$ and the latter $K=2$.
The visualization of the corresponding atoms and reconstructed patches in Figures~\ref{fig:MSEcameraman} and~\ref{fig:MSEnet} (b-f) shows that R-K-SVD better preserves edge sharpness as no averaging over rotations is happening (see also Fig.~\ref{fig:toyCoding}).
We also observed that a standard K-SVD yields atoms that are approximate rotations of each other (e.g., Fig.~\ref{fig:MSEcameraman} (e)),
meaning that the atom budget is spent on representing rotations of unique patterns.

In our texture classification experiment,
we show that a simple classifier based on rotational sparse coding
can achieve perfect performance on the Outex\_TC\_00010 dataset.
While modern CNNs can approach this result,
they must be pretrained and are not as easy to interpret
as the rotational sparse coding model.
Our approach is reminiscent of (and partially inspired by) textons~\cite{Malik_Contour_2001},
with the difference that, here, our textons are not clusters in a feature space,
but rather entire dictionaries that encode complicated surfaces.
However, our aim is not to promote this specific classifier.
Instead, we think the texture classification experiments
illustrate a key feature of the rotational formulation of dictionary learning:
the tradeoff between representation and discrimination.
Because the rotational dictionaries have a better model for the formation of patches,
they can lie closer to the true data,
allowing accurate representations without a loss of discrimination
(i.e., an explosion of the span as discussed with Figure~\ref{fig:toyCoding}).
Certainly, tackling challenges like material classification or
even texture classification with more intraclass variation,
e.g., variations in scale,
will require more complicated classification schemes than we discussed here.
We see potential in supervised and discriminative sparse coding formulations
for learning the atoms, such as those proposed in~\cite{MPB2008,jiang_label_2013,bahrampour_multimodal_2016}.
In addition, it may be fruitful to combine the rotational formulation,
which provides hand-designed invariances,
with deep neural networks,
which can learn very effective image features,
along the lines of the Deep TEN architecture~\cite{zhang_deep_2017}
or the rotation invariant CNN in \cite{marcos_learning_2016}.

\section*{Acknowledgments}
We thank Mikkel Schmidt for their willingness to provide additional details
regarding the speed of the method presented in~\cite{morup_transformation_2011}.

\appendices

\begin{table*}[!htbp]
  \centering%
  \begin{threeparttable}
    \caption{Texture classification training cross validation accuracy$^{1,2}$}%
    \begin{tabular}{rrrrcrrrcrrrcrrr}\toprule
      &  \multicolumn{3}{c}{$M = 10$} && \multicolumn{3}{c}{$M = 25$} && \multicolumn{3}{c}{$M = 50$} && \multicolumn{3}{c}{$M = 100$} \\
      \cmidrule{2-4} \cmidrule{6-8} \cmidrule{10-12} \cmidrule{14-16} 
      $ N = $&   $11$ &  $13$ &  $15$ & &  $11$ &  $13$ &  $15$ & &  $11$ &  $13$ &  $15$ & &  $11$ &  $13$ &  $15$ \\\midrule
      {\bf K-SVD} &&&&&&&&&&&&&&& \\
      $K = 1$ & 0.990 & 0.988 & 0.984 && 0.993 & 0.992 & 0.992 && 0.994 & 0.995 & 0.996 && 0.996 & 0.997 & 0.998 \\
      $K = 2$ & 0.987 & 0.988 & 0.981 && 0.995 & 0.993 & 0.993 && 0.995 & 0.995 & 0.996 && 0.997 & 0.998 & 0.997 \\
      $K = 3$ & 0.982 & 0.984 & 0.984 && 0.993 & 0.993 & 0.995 && 0.997 & 0.996 & 0.996 && {\bf 0.999} & 0.998 & 0.998 \\
      {\bf aug-K-SVD} &&&&&&&&&&&&&&& \\
      $K = 1$ & 0.972 & 0.968 & 0.975 && 0.979 & 0.985 & 0.987 && 0.988 & 0.989 & 0.991 && 0.989 & 0.991 & {\bf 0.993} \\
      $K = 2$ & 0.967 & 0.968 & 0.968 && 0.981 & 0.987 & 0.992 && 0.985 & 0.991 & 0.992 && 0.987 & 0.991 & 0.992 \\
      $K = 3$ & 0.971 & 0.964 & 0.961 && 0.984 & 0.990 & 0.987 && 0.986 & 0.989 & 0.990 && 0.989 & 0.991 & 0.992 \\
      {\bf R-K-SVD} &&&&&&&&&&&&&&& \\
      $K = 1$ & 0.969 & 0.977 & 0.979 && 0.975 & 0.980 & 0.985 && 0.982 & 0.985 & 0.988 && 0.983 & 0.988 & {\bf 0.990} \\
      $K = 2$ & 0.963 & 0.966 & 0.978 && 0.977 & 0.979 & 0.984 && 0.979 & 0.982 & 0.986 && 0.981 & 0.986 & 0.987 \\
      $K = 3$ & 0.954 & 0.966 & 0.968 && 0.968 & 0.979 & 0.982 && 0.973 & 0.982 & 0.986 && 0.977 & 0.987 & 0.990 \\
      \bottomrule
    \end{tabular}%
    \label{tab:tex-param}%
    \begin{tablenotes}
      \item {\footnotesize $^1$Values are from a hold-out cross validation on the training data only:
          the image-level classifier is trained using 12\% of the training data (selected at random),
          and the remaining 88\% is used to compute accuracy.
          The process is repeated 100 times and the results are averaged. }%
          \item {\footnotesize $^2$$M$ is dictionary size, $N$ is patch size, and $K$ is sparsity level.}
    \end{tablenotes}
  \end{threeparttable}
\end{table*}

\section{Cross-Validation Results}
\label{app:sweep}
Table~\ref{tab:tex-param} gives the full results of the parameter sweep
described in Section~\ref{sec:texture}.\textbf{Classifier}
and discussed in Section~\ref{sec:texture}.\textbf{Results}.

\section{Deep Learning Comparison Details and Full Results}\label{app:deep_results}
We compare the results obtained on the Outex\_TC\_00010 dataset with various state of the art CNN architectures, namely VGG16, VGG19 \cite{SiZ2014}, ResNet50 \cite{he2016deep}, DenseNet121, DenseNet169, DenseNet201 \cite{HLW2017} and InceptionV3 \cite{szegedy2016rethinking}.
The networks are either trained from \textit{scratch} (random weights initialization) or \textit{finetuned} using weights pre-trained on ImageNet \cite{imagenet_cvpr09}.
In both training schemes, the last layer of the networks is replaced by a randomly initialized dense softmax layer with 24 neurons. 
The Outex images are pre-processed in the same manner as the ImageNet images used for pre-training the networks (generally zero mean and unit variance using ImageNet training set statistics).
Note that the input images have to be resized to $224\times224$ for training and evaluating the VGG16 and VGG19 as these networks require fixed input sizes (no global average pooling).
The networks are trained for 100 epochs using an Adam optimizer with categorical cross-entropy and standard hyper-parameters:  $\beta_1 = 0.9$, $\beta_2 = 0.999$, batch size $= 32$, learning rate $= 10^{-4}$, decay rate $= 10^{-3}$. 
The data augmentation employed in these experiments includes random vertical and horizontal flips as well as random rotations in the interval $[0,\pi]$.
We used a Keras implementation with TensorFlow backend and trained and evaluated the networks with a Titan Xp graphic card.

Resulting classification performances are reported in Table~\ref{tab:res_deep}.
The best result (99.82\% accuracy) is achieved with a pre-trained DenseNet121,  finetuned and using data augmentation. 
The top three results are also reported alongside the main results in Table~\ref{tab:tex-results}.

\begin{table}[h]
  \centering
  \caption{Average accuracy and standard deviation (10 repetitions)}
\begin{tabular}{lc}\toprule
  network and training & accuracy  \\
\midrule
  VGG16 scratch & 0.4233  {$ \pm {0.0501}$} \\
  VGG16 scratch + data-augm. & 0.8968 {$ \pm {0.0331}$} \\
  VGG16 pre-trained & 0.5740 {$ \pm {0.0290}$} \\
  VGG16 pre-trained + data-augm. & 0.9850 {$ \pm {0.0067}$} \\
  \hline 
  VGG19 scratch & 0.3198 {$ \pm {0.0520}$} \\
  VGG19 scratch + data-augm. & 0.8431 {$ \pm {0.0672}$} \\
  VGG19 pre-trained & 0.5966 {$ \pm {0.0347}$} \\
  VGG19 pre-trained + data-augm. & 0.9790 {$ \pm {0.0074}$} \\
  \hline 
  ResNet50 scratch & 0.3930 {$ \pm {0.0413}$} \\
  ResNet50 scratch + data-augm. & 0.6875 {$ \pm {0.0635}$} \\
  ResNet50 pre-trained & 0.8807 {$ \pm {0.0203}$} \\
  ResNet50 pre-trained + data-augm. & 0.9945 {$ \pm {0.0041}$} \\
  \hline 
  DenseNet121 scratch & 0.6308 {$ \pm {0.0187}$} \\
  DenseNet121 scratch + data-augm. & 0.9611 {$ \pm {0.0158}$} \\
  DenseNet121 pre-trained & 0.9034 {$ \pm {0.0102}$} \\
  DenseNet121 pre-trained + data-augm. & 0.9982 {$ \pm {0.0016}$} \\
  \hline 
  DenseNet169 scratch & 0.6520 {$ \pm {0.0130}$} \\
  DenseNet169 scratch + data-augm. & 0.9501 {$ \pm {0.0297}$} \\
  DenseNet169 pre-trained & 0.9252 {$ \pm {0.0116}$} \\
  DenseNet169 pre-trained + data-augm. & \textbf{0.9983 {$ \pm {0.0010}$}} \\
  \hline 
  DenseNet201 scratch & 0.6450 {$ \pm {0.0202}$} \\
  DenseNet201 scratch + data-augm. & 0.9621 {$ \pm {0.0140}$} \\
  DenseNet201 pre-trained & 0.9401 {$ \pm {0.0092}$} \\
  DenseNet201 pre-trained + data-augm. & 0.9980 {$ \pm {0.0011}$} \\
  \hline 
  InceptionV3 scratch & 0.6323 {$ \pm {0.0347}$} \\
  InceptionV3 scratch + data-augm. & 0.8479 {$ \pm {0.0324}$} \\
  InceptionV3 pre-trained & 0.8888 {$ \pm {0.0248}$} \\
  InceptionV3 pre-trained + data-augm. & 0.9944 {$ \pm {0.0018}$}\\\bottomrule
\end{tabular}
\label{tab:res_deep}
\end{table}
 
\bibliographystyle{IEEEtran}
\bibliography{refsAdrien,refsMike,refsVincent,refsExtra}

\begin{IEEEbiography}{Michael McCann} (S'10-M'15) received the B.S.E. in biomedical engineering in 2010 from the University of Michigan
  and the Ph.D. degree in biomedical engineering from Carnegie Mellon University in 2015.
  From 2015-2019, he worked with Michael Unser's group at the \'{E}cole polytechnique f\'{e}d\'{e}rale de Lausanne (EPFL).
  He is currently a Postdoctoral Research Associate with Saiprasad Ravishankar's group
  in the Dept. of Computational Mathematics, Science and Engineering
  at Michigan State University.
  His research interests center on developing algorithms for biomedical image reconstruction
  using tools from signal processing, optimization, and machine learning.
\end{IEEEbiography}%

\begin{IEEEbiography}
{Vincent Andrearczyk}
Vincent Andrearczyk received a double Masters degree in electronics and signal processing from ENSEEIHT, France and Dublin City University, in 2012 and 2013 respectively.
He completed his PhD degree on deep learning for texture and dynamic texture analysis at Dublin City University in 2017.
He is currently a post-doctoral researcher at the University of Applied Sciences and Arts Western Switzerland with a research focus on deep learning for medical image analysis.
\end{IEEEbiography}

\begin{IEEEbiography}
{Michael Unser} (M'89-SM'94-F'99) is professor and director of EPFL's Biomedical Imaging Group, Lausanne, Switzerland.
His primary area of investigation is biomedical image processing.
He is internationally recognized for his research contributions to sampling theory, wavelets, the use of splines for image processing, stochastic processes, and computational bioimaging.
He has published over 250 journal papers on those topics.
He is the author with P. Tafti of the book \textit{An introduction to sparse stochastic processes}, Cambridge University Press 2014. From 1985 to 1997, he was with the Biomedical Engineering and Instrumentation Program, National Institutes of Health, Bethesda USA, conducting research on bioimaging. Dr. Unser has held the position of associate Editor-in-Chief (2003-2005) for the IEEE Transactions on Medical Imaging. He is currently member of the editorial boards of SIAM J. Imaging Sciences, IEEE J. Selected Topics in Signal Processing, and Foundations and Trends in Signal Processing. He is the founding chair of the technical committee on Bio Imaging and Signal Processing (BISP) of the IEEE Signal Processing Society. Prof. Unser is a fellow of the IEEE (1999), an EURASIP fellow (2009), and a member of the Swiss Academy of Engineering Sciences. He is the recipient of several international prizes including three IEEE-SPS Best Paper Awards and two Technical Achievement Awards from the IEEE (2008 SPS and EMBS 2010).
\end{IEEEbiography}

\begin{IEEEbiography}
{Adrien Depeursinge} (M'06)
Adrien Depeursinge received the B.Sc. and M.Sc. degrees in electrical engineering from the Swiss Federal Institute of Technology (EPFL), Lausanne, Switzerland with a specialization in signal processing. From 2006 to 2010, he performed his Ph.D. thesis on medical image analysis at the University Hospitals of Geneva (HUG). He then spent two years as a Postdoctoral Fellow at the Department of Radiology of the School of Medicine at Stanford University. He has currently a joint position as an Associate Professor at the Institute of Information Systems, University of Applied Sciences Western Switzerland (HES-SO), and as a Senior Research Scientist at EPFL.
\end{IEEEbiography}

\end{document}